\documentclass[10pt, journal, twocolumn]{IEEEtran}  
\usepackage{amssymb,amsmath}
\usepackage{cite}
\usepackage{graphicx,subfigure}
\usepackage{psfrag}
\usepackage{url}
\usepackage{hyperref}
\usepackage{booktabs}
\usepackage[latin1]{inputenc}
\usepackage[absolute,overlay]{textpos}
\usepackage[latin1]{inputenc}
\usepackage{color}
\usepackage{colortbl}
\definecolor{lightgray}{gray}{0.95}
\definecolor{lightgray2}{gray}{0.8}
\usepackage{multirow}

\usepackage{array}
\newcolumntype{L}[1]{>{\raggedright\let\newline\\\arraybackslash\hspace{0pt}}m{#1}}
\newcolumntype{R}[1]{>{\raggedleft\let\newline\\\arraybackslash\hspace{0pt}}m{#1}}
\usepackage{arydshln}

\begin{document}
	
	\title{High-Power and Safe RF Wireless Charging: \\
  Cautious Deployment and Operation}
	\author{
		\IEEEauthorblockN{
		Onel L. A. L\'opez, 
		Osmel M. Rosabal,
		Amirhossein  Azarbahram,
            A. Basit Khattak,
            Mehdi Monemi,\\
            Richard D. Souza, 
            Petar Popovski, and
            Matti Latva-aho
		}
	\thanks{Onel L. A. L\'opez, Osmel M. Rosabal, Amirhossein Azarbahram, A. Basit Khattak, Mehdi Monemi, and  Matti Latva-aho are with the Centre for Wireless Communications (CWC), University of Oulu, Finland. \{onel.alcarazlopez, osmel.martinezrosabal, amirhossein.azarbahram, abdul.khattak, mehdi.monemi,  matti.latva-aho\}@oulu.fi.} 
 \thanks{Richard Souza is with the Federal University of Santa Catarina, Florian\'opolis, SC, Brazil. richard.demo@ufsc.br.} 
 \thanks{Petar Popovski is with the Department of Electronic Systems, Aalborg University, Aalborg, Denmark. petarp@es.aau.dk.} 
	\thanks{This work is partially supported by the Finnish Foundation for Technology Promotion, Academy of Finland (Grants 348515 and 346208 (6G Flagship)), the European Commission through the Horizon Europe/JU SNS project Hexa-X-II (Grant Agreement no. 101095759), CNPq/Brazil (402378/2021-0, 305021/2021-4, 401730/2022-0) and RNP/MCTI/Brazil 6G Mobile Communications Systems (01245.010604/2020-14).}}
	\maketitle
	\begin{abstract}
		%
	The wired charging and the need for battery replacements are critical barriers to unlimited, scalable, and sustainable mobile connectivity, motivating the interest in radio frequency (RF) wireless power transfer (WPT) technology. However, the inherently low end-to-end power transfer efficiency (PTE) and health/safety-related apprehensions about the technology are critical obstacles. Indeed, RF-WPT implementation and operation require efficient and cautious strategies and protocols, especially when targeting high-power charging, which constitutes the scope of this work. Herein, we overview the main factors affecting the end-to-end PTE of RF-WPT systems and their multiplicative effect and interdependencies. Moreover, we discuss key electromagnetic field (EMF) exposure metrics, safety limits, and approaches for efficient and EMF-aware deployment and operation. Quantitatively, we show that near-field RF charging may significantly reduce EMF exposure, and thus must be promoted. We also present our vision of a cyber-physical system for efficient and safe wireless charging, specify key components and their interrelation, and illustrate numerically the PTE attained by two modern low-power multi-antenna architectures in a simple setup. Throughout the paper, we highlight the need for high end-to-end PTE architectures and charging protocols transparently complying with EMF exposure regulations and outline relevant challenges and research directions. This work expands the vision and understanding of modern RF-WPT technology and constitutes a step towards making the technology attractive for worldwide commercial exploitation.
	\end{abstract}


	\section{Introduction}\label{intro}
The rise of wireless communications leads to an increasingly digitized, data-driven, autonomous society. Unfortunately, the wired charging and the need for battery replacements are still critical barriers to unlimited, scalable, and sustainable mobile connectivity. Ubiquitous wired charging is usually cost-prohibitive, especially in industrial Internet of Things (IoT) deployments, while battery-powered solutions struggle with limited battery life and the replacement problem, which is not economical nor eco-friendly. 
Industry and academia agree that ambient energy harvesting (EH)
and wireless power transfer (WPT) paradigms may break down the previous barrier shortly \cite{Lopez.2021}. Notably, WPT may be inevitably needed in applications with quality-of-service (QoS) requirements since EH from ambient energy is non-deterministic and location-dependent, often requiring relatively large form factor transducers. 

WPT solutions widely available in the market rely on short-range reactive techniques exploiting magnetic or electric fields with limited multi-user support and still confining the maneuverability, mobility, and scalability of the device(s) being charged. 
Conversely, radiative radio frequency (RF)-WPT technology natively supports multiuser charging 
and allows greater charging radii \cite{Lopez.2021}. However,  the inherently low end-to-end power transfer efficiency (PTE) of RF wireless charging and safety-related apprehensions 
are critical obstacles, slowing down standardization activities.
These have motivated mostly customized low-power IoT charging applications while relying on i) energy beamforming (EBF) and waveform optimization \cite{Ginting.2020,Clerckx.2021,Lopez.2022,Zhang.2022,Azarbahram.2023}; ii) distributed and massive antenna systems \cite{Lopez.2022}; iii) smart reflect arrays and reconfigurable metasurfaces \cite{Zhou.2022,Azarbahram.2023}; and iv) flexible energy transmitters (ETs) \cite{Lopez.2021} (i.e., moving ETs and ETs equipped with rotary antennas); while in 
many cases explicitly incorporating electromagnetic field (EMF) radiation exposure control~\cite{Ginting.2020,Lopez.2022}.
%
\begin{figure*}[t!]
    \centering
    \includegraphics[width=0.93\linewidth]{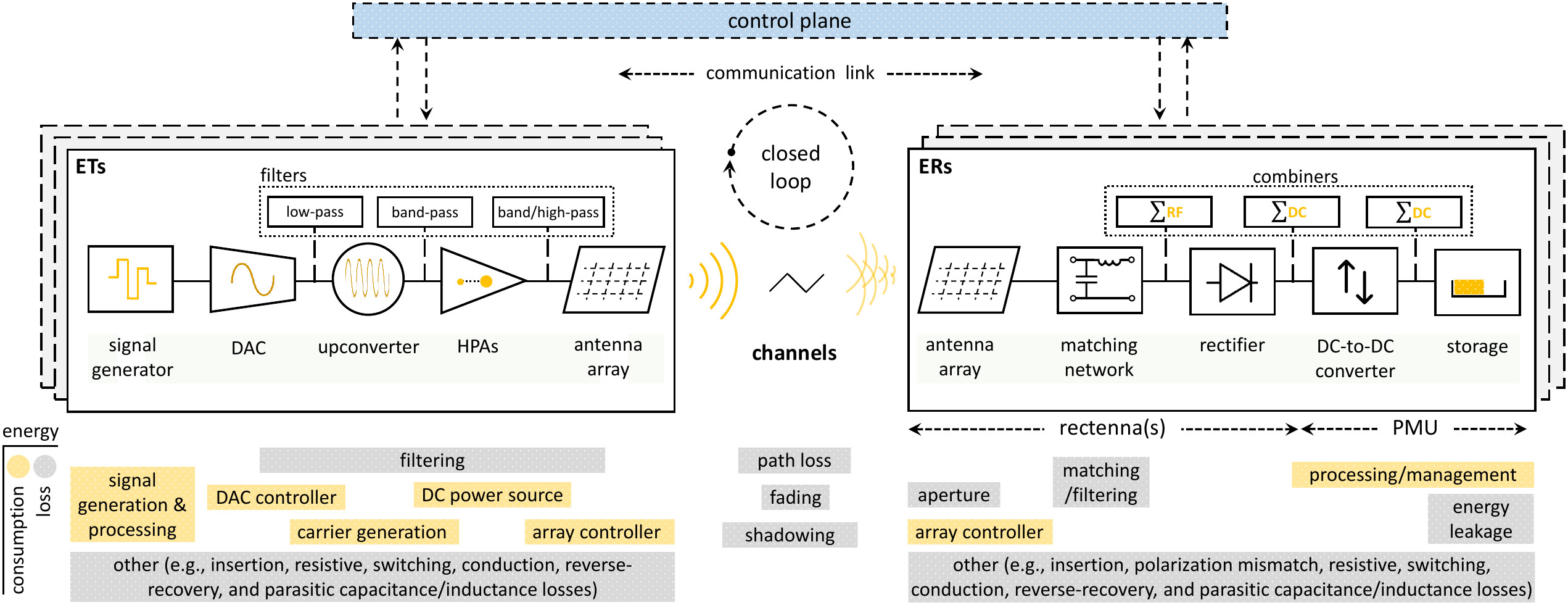}
    \caption{Block diagram of the architecture of an RF-WPT system and main energy consumption/loss sources.}
    \label{fig:system}
\end{figure*}
We believe that a comprehensive combination and holistic optimization of these promising technologies (and others) are needed to make the PTE of RF-WPT systems competitive, expanding the market to relatively high-power charging applications. Indeed, some startups
such as \href{https://energous.com}{Energous}, \href{https://ossia.com}{Ossia} and \href{https://guru.inc}{GuRu} are taking a leap forward with a few RF-WPT solutions that are not restricted to this regime.

Although RF-WPT is rapidly gaining market, its implementation and operation still require 
cautious strategies and protocols. Herein we elaborate further on this while making the following 
contributions: 
\begin{itemize}
    \item We overview the main factors affecting the PTE of RF-WPT systems,
    as well as their interrelation, which calls for joint optimization approaches.
    \item We overview key EMF exposure metrics and safety limits, and discuss approaches for 
   EMF-aware operation. We illustrate the effect of the frequency and far/near-field conditions on EMF exposure manageability,
   showing that near-field RF charging may significantly reduce EMF exposure, and thus must be promoted. 
    \item We present our vision of a cyber-physical system for efficient and safe WPT while specifying key components and their interrelation. We illustrate 
    the PTE attained by two modern low-power MIMO architectures, showing that intelligent surface (RIS) and dynamic metasurface antenna (DMA) based ETs may attain similar performance, also in terms of EMF radiation manageability.
    \item We highlight relevant challenges throughout the paper while pointing to potential research directions.
\end{itemize}
%

\emph{Reproducible research:} The simulation code is available at: \href{https://github.com/onel2428/High_Power_and_Safe_WPT.git}{github.com/onel2428/High\_Power\_and\_Safe\_WPT.git}.

%
%
\section{RF-WPT Technology and Key Concerns} 
 The architecture of an RF-WPT system is illustrated in Fig.~\ref{fig:system} and comprises 
 i) ETs; ii) wireless channels; and iii) energy receivers (ERs). On the ET side, the charging signal is generated in the digital domain and converted to analog by the digital-to-analog converter (DAC). This is followed by frequency upconversion to the RF domain, signal amplification supported by a high-power amplifier (HPA), and transmission over the wireless medium by one or multiple antennas. On the ER side, the impinging charging signal, distorted by the channel, is captured by the receive antenna(s) and converted to DC via a rectifying network, all composing the so-called rectenna(s). 
 A matching network between the antenna(s) and the rectifying circuitry is needed to maximize the power transfer at a designed frequency band. 
 The rectified signal goes through a power management unit (PMU), which may include a DC-to-DC converter and energy storage,  interfacing with the energy demands of the device.
 %


 Next, we discuss the two main concerns about RF-WPT technology.  Addressing them effectively is crucial for enhancing the technology's feasibility and attractiveness while propitiating more standardization attempts, which in turn guarantee interoperability 
 and expand market accessibility.



%
\subsection{Low End-to-End PTE}\label{pte}
  Charging based on RF-WPT is slower and less efficient than traditional wired and reactive wireless charging technologies. This is due to the many sources of energy consumption and loss in the system as depicted in Fig.~\ref{fig:system}. 
 
  \begin{table*}[t!]
    \caption{EMF exposure limits for the general public at  high frequencies  according to ICNIRP}
    \label{tab:EMF}
     \centering
    \begin{tabular}{ p{3.6cm}  p{1.4cm} p{1.3cm} p{3.5cm} p{4.7cm} }
        \toprule      
       \multirow{2}{*}{\textbf{metric}} & \textbf{exposure} & \textbf{averaging}  & \multicolumn{2}{c}{\textbf{frequency range}}\qquad\qquad  \\
       \cline{4-5}  & \textbf{zone}  & \textbf{time (min)} & $2-6$ GHz & $6-300$ GHz   \\
         \midrule
      \multirow{2}{*}{incident power density (W/$\mathrm{m}^2$)} & whole body & 30   & 10 & 10   \\
         & local & 6 & 40 & $55/f(\text{GHz})^{0.177}$  \\     
      \cline{1-5} incident energy density (kJ/$\mathrm{m}^2$) & local & $t (t<6)$ & $14.4\times (0.05+0.95\sqrt{t/6})$ & $19.8\times (0.05+0.95\sqrt{t/6})/f(\text{GHz})^{0.177}$    \\
         \bottomrule
    \end{tabular}     
\end{table*}
  
  The main energy consumption sources are the baseband processing, including the digital signal generation and DAC controller, the carrier generation, and the HPA DC supply at the ET, while the signal processing and energy management may dominate on the ER side. Additional energy may be consumed by the antenna array controlling circuitry. 
  
  The main energy losses come from filtering processes, channel impairments, and the receive antenna aperture. Filtering may be applied i) after the DAC, to remove high-frequency components; ii) after upconversion, to remove parasite frequency components; iii) after the HPA, to remove frequency components introduced by the HPA non-linearity; iv) in the matching network, since frequency components outside its operation bandwidth are removed or attenuated; and v) after the rectifier, to remove the fundamental and harmonic frequencies of the rectified signal. Meanwhile, the channel impairments encompass path loss, fading, shadowing, and non-line-of-sight conditions, and cause the greatest efficiency loss in typical systems. 
Other energy losses may come from i) elements insertion in the case of not-integrated circuits; ii) transmit-receive polarization mismatches; 
iii) energy leakage in the energy storage element; and iv) energy dissipation at resistive elements, switching, conduction, reverse-recovery, and parasitic capacitance/inductance.

Notice that the efficiency of each block in the RF-WPT system
has a multiplicative effect on the end-to-end PTE. The performance of the blocks may exhibit some dependence, so optimizing them independently is not optimal. For instance:
\begin{itemize}
    \item DACs consume less energy as the required resolution decreases. However, lower DAC resolutions trigger the generation of high-frequency components that must be filtered out, leading to energy loss;
    \item In multi-tone charging waveforms, the more subcarriers, the greater the flexibility to optimize the waveform to drive the HPA and rectenna to high-efficiency regimes \cite{Clerckx.2021}, but the greater the signal processing energy consumption;
    \item Reducing the number of transmit RF chains is appealing for reducing the transmit energy consumption, but might also limit the EBF gains of the transmit antenna array and the capability to charge multiple users simultaneously.
\end{itemize}
These fundamental trade-offs must be properly addressed at the design and operation phase. 
Regarding the latter, an efficient high-power RF-WPT system must operate in a closed loop. At least a reverse or forward communication link is needed to acquire channel state information (CSI) and combat channel impairments~\cite{Clerckx.2021,Ginting.2020,Lopez.2022,Zhang.2022}.
More evolved optimizations require forward and reverse communication links supporting a control plane. 


\subsection{EMF Exposure Concerns}  \label{fear}  
%
 RF transmissions are non-ionizing, thus unable to directly alter the structure of molecules, which would lead to carcinogenic effects. However, this does not discard potential indirect links. The International Agency for Research on Cancer (IARC)
  classified RF emissions as ``possibly carcinogenic to humans'' \cite{iarcclassification}.\footnote{This classification corresponds to Group 2B, while Group 1, 2A, and 3 correspond to agents that are ``carcinogenic'', ``probably carcinogenic'', and ``not classifiable as to its carcinogenicity'' to humans \cite{iarcclassification}. }  This classification does not indicate that RF emissions definitively cause cancer, but that there is (limited) evidence suggesting a possible link. 
  Nevertheless, it is well known that mild biological effects such as tissue heating may occur when exposed to very strong EMFs at microwave frequencies or higher.  
 
EMF exposure limits have been set by regulatory bodies, such as the United States Federal Communications Commission (FCC) and Food \& Drug Administration (FDA), and the International Telecommunication Union (ITU), to mitigate the potential adverse effects on human health. The regulations are intended to balance the potential risks and benefits of using RF technologies and are regularly reviewed and updated. 
There are also standard bodies, such as the American National Standards Institute (ANSI), the Institute of Electrical and Electronics Engineers (IEEE), and the International Commission on Non-Ionizing Radiation Protection (ICNIRP),  which can not legally enforce their recommended limits or protocols but whose specifications are usually adopted by regulators.

EMF exposure metrics such as specific absorption rate (SAR, in W/kg), specific absorption (SA, in kJ/kg), absorbed power density (APD, in W/m$^2$), and absorbed energy density (AED, in kJ/m$^2$) correspond to physical quantities that are closely related to RF-induced biological effects. SAR and SA are valid for systems operating below 6~GHz, where EMFs penetrate deep into the tissue, while APD and AED are more useful above 6~GHz where penetration depth is less relevant \cite{ICNIRP.2020}.
However, they may not be easily measured/estimated in practice. Other related quantities that are more easily evaluated are the incident power density (in W/m$^2$), incident energy density (kJ/m$^2$), electric field strength (in V/m), and magnetic field strength (in A/m). EMF exposure limits are usually conservatively set in terms of them such that SAR, SA, APD, and AED values remain below the safety limits with extremely high reliability \cite{ICNIRP.2020}. 


High-power RF charging may be possible only for short ranges as the channel losses increase significantly with the distance. 
In short-range scenarios, precise (highly focused) spatial charging is needed to avoid increased EMF exposure levels in the surroundings of the ERs. This may be only achieved by operating at relatively large frequencies and in the massive MIMO regime \cite{Lopez.2022}, but also in the near-field region of the ETs \cite{Zhang.2022}. 

Table~\ref{tab:EMF} compiles the EMF exposure limits for the general public according to ICNIRP.\footnote{Europe and FCC mostly follow ICNIRP recommendations for EMF exposure limits. The limits illustrated in Table~\ref{tab:EMF} for occupational scenarios, where individuals are trained to be aware of potential EMF risks and to employ appropriate harm-mitigation measures, are five-fold less stringent than for the general public.} The EMF limit values decrease slowly with the operation frequency above 6~GHz in both incident power and energy densities. Moreover, the incident power density limit is far more stringent for local exposure zones than for the whole body, for which the averaging time is also greater.

\begin{figure}[t!]
    \centering
    \includegraphics[width=0.95\linewidth]{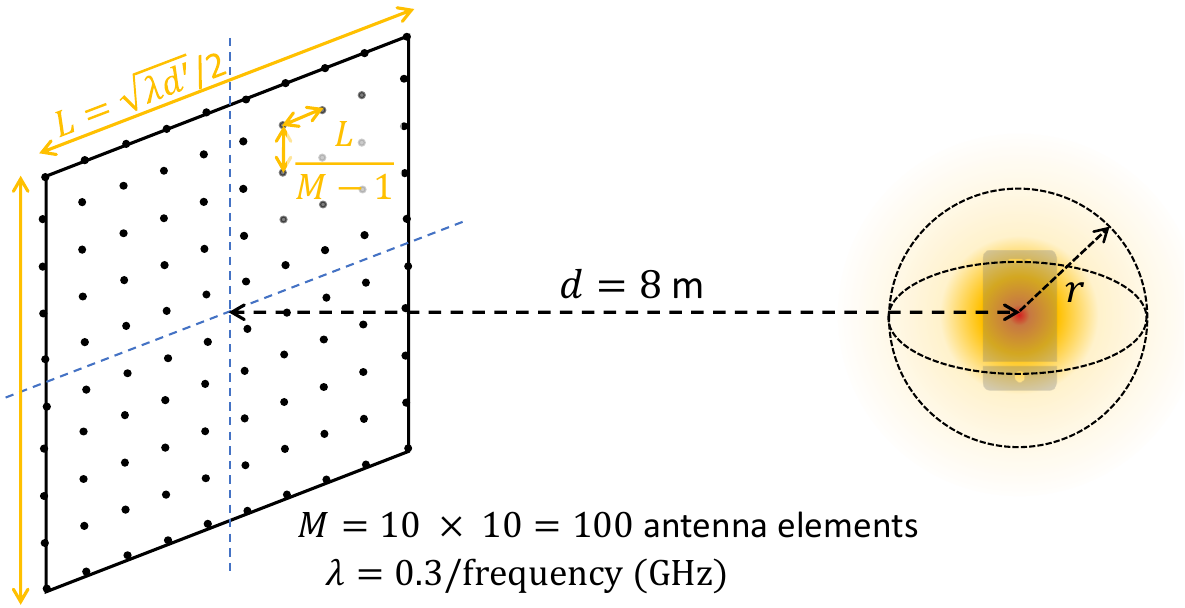}\\
    \vspace{4mm}
    \includegraphics[width=\linewidth]{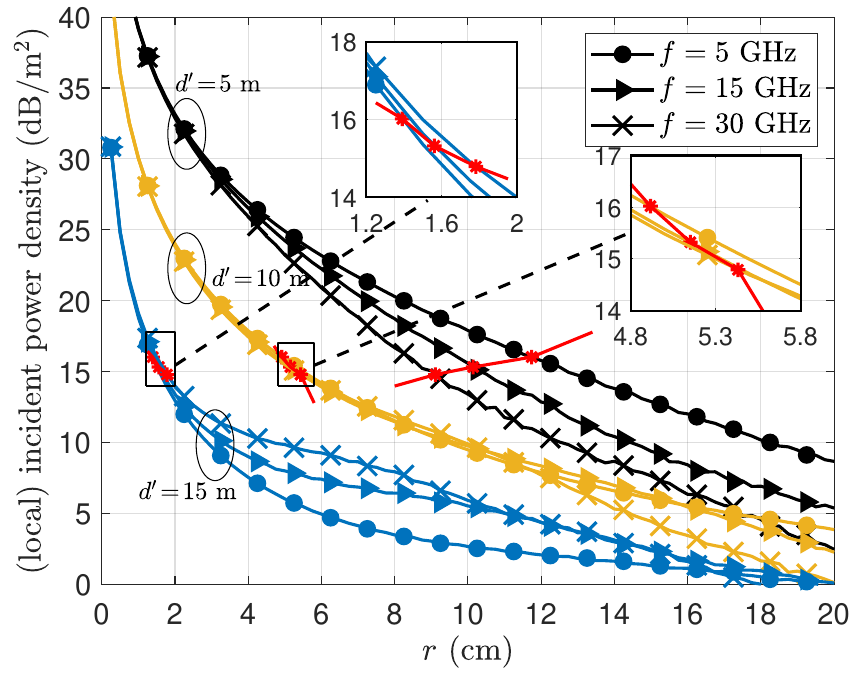}
    \caption{a) Simulation setup: a fully digital transmit antenna array composed of $10\times 10$ antenna elements charges 
    a single-antenna ER located 8 meters away 
    in its boresight direction. b) Incident RF power density at a distance $r$ from the ER and normalized by the RF power delivered to the device. Results are presented for three operation frequencies and near-far-field thresholds $d'$ considering purely geometrical channels and antenna elements with boresight gains of 13 dB. EMF limits for local exposure according to ICNIRP values are illustrated by red lines and $*$ markers. 
    }
    \label{fig:result1}
\end{figure}

Consider the setup illustrated in Fig.~\ref{fig:result1}a, where a $10\times 10$ antenna array is charging a device $d=8$ m away in its boresight direction. The array is square with horizontally/vertically equally-spaced antenna elements and $L$ denotes the edge length, which we allow to scale as a function of the wavelength $\lambda$ and the near-far-field threshold distance $d'$, where $d'=L^2/\lambda$ is a common threshold between near and far field conditions, i.e., $d< d'\ (d> d')\rightarrow$ near (far) field. 
In Fig.~\ref{fig:result1}b, we show the incident power density at a distance $r$ from the receive antenna, i.e., in the $r-$radius sphere centered at the receive antenna position, and normalized by the RF power delivered to the device, while considering purely geometrical channels. High EMF exposure levels may come from potentially high spatial correlations,
which decrease with $r$. Therefore, the incident power density decreases with $r$ (an observation also made in \cite{Lopez.2022}), especially under near-field conditions (large $d'$). It must be noted that under far-field conditions strictly directional beams are created,
and thus the local EMF exposure is significantly greater if the living tissues are between the antenna array and the ER, and one must rely mostly on beam avoidance mechanisms \cite{Ginting.2020}. This is not a problem as near-field conditions intensify, by decreasing $d/d'$, since the RF energy reaches the receiver from a broader array of spatial directions and is sharply focused on the receive antenna. 

The impact of frequency, in Fig.~\ref{fig:result1}b, is significantly less obvious: given a fixed number of antennas, the EMF exposure can be more easily controlled at relatively high (low) frequencies when operating in far (near)-field conditions. If the number of antennas is allowed to scale with the frequency (e.g., assuming half-wavelength antenna spacing), the spatial resolution brought by the number of antennas at high frequencies will always be preferred to easily control the EMF exposure \cite{Lopez.2022}. However, infinitely increasing the number of antennas in a small-scale setup may not be efficient, which motivates the setup and results illustrated here. We have also delimited in Fig.~\ref{fig:result1}b the local exposure limits as set by ICNIRP (Table~\ref{tab:EMF}) assuming 1 W of RF charging power at the device and uninterrupted transmissions such that the results coming from 6-mins averaging match the instantaneous results. Notably, in very near-field conditions such that $d'\ge 15$~m, the incident EMF exposure is below the safety limits for $r\le 1.8$~cm and $f\le 30$ GHz, which might be enough for safely charging a phone in human hands if the receive antenna and phone are properly placed and hold.

\begin{figure*}[t!]
    \centering
    \includegraphics[width=0.85\linewidth]{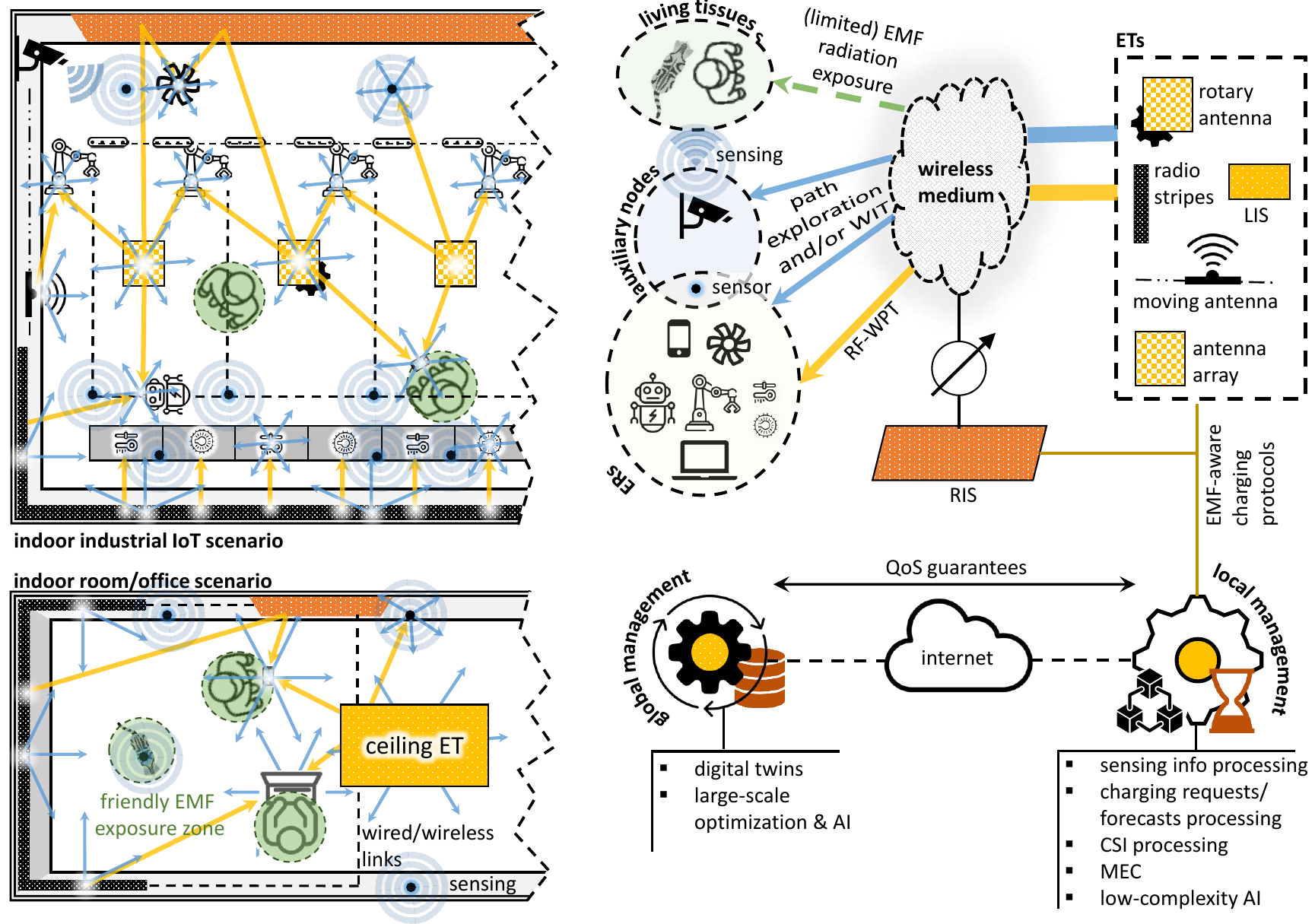}
    \caption{Our vision for a competitive RF-WPT system and two key scenarios: room/office and industrial IoT. Charging arrows are only indicative of the transmission origin since charging may be highly spatial-focused when operating in the near field.}
    \label{fig:vision}
\end{figure*}

The above results are for local incident power density. Getting quantitative insights on the whole body EMF exposure, as for a person holding the device, may be much more intricate as it depends on the body's build. A child holding the ER would have much more EMF exposure than an adult as the entire body is closer to the device. How to estimate the incident power density for the whole body in near-field conditions is open for research, and conservative methodologies are in general preferred. It might be easier to compute volumetric, rather than area, power/energy densities, but this either requires some research on corresponding EMF exposure limit values or proper procedures for quantities conversions. Regarding local incident energy density, one must note that this EMF exposure constraint is set to avoid high peaks of power densities for up to several minutes that would still meet the incident power density constraint. If, on the contrary, the instantaneous incident power density remains below or equal to the 6-min average value, the incident energy density constraint will be always met. In any case, the 6-min averaging of power and energy densities gives some flexibility to potential charging protocols and thus must be certainly explored with dedicated research.
Anyway, cautious strategies for the deployment and operation of RF-WPT systems are needed. They must provide transparent EMF radiation exposure guarantees for the end users.

\section{A Cyber-physical System for Efficient and Safe Wireless Charging}\label{cps_system}


The key components and vision for a competitive indoor WPT cyber-physical system for two key scenarios are illustrated in Fig.~\ref{fig:vision}. 
We believe that a proper holistic design of these components, although extremely challenging, is critical to unleashing the full potential of RF-WPT and turning it into a sustainable wireless charging technology, bringing the attention of the market, propitiating more standardization attempts and commercial products.

	\subsection{Transmitters and Intelligent Reflectors}\label{ETs}

	
Massive MIMO deployments can compensate for the extremely large channel losses, promote near-field conditions, and support ubiquitous energy access \cite{Lopez.2021,Lopez.2022}. However, the required high manufacturing and operating expenses, as well as the increased power consumption, make the development and deployment of truly large-scale antenna arrays challenging, motivating the research on more affordable and low-power MIMO architectures, potentially powered by ambient EH sources, that can scale with the number of antennas more sustainably \cite{Lopez.2023}. 

Hybrid analog-digital MIMO implementations require a reduced number of power-consuming RF chains, while further power reductions can be achieved by using low-resolution DACs. In the analog processing domain, one can rely on antenna selection architectures, 
parasitic, load-modulated, and lens arrays.
Metasurface-aided architectures, as RIS and large intelligent surface (LIS), constitute another key research front to support low-power MIMO \cite{Long.2021,Zhou.2022,Azarbahram.2023}. RISs can efficiently assist already-deployed MIMO networks by on-the-fly and opportunistic  ``reconfiguration'' of the wireless medium, directing the power towards the intended ER(s), by means of dynamically adapting phase, amplitude, frequency, and polarization of transmit signals.
 Conventional RISs are composed of fully passive reflecting elements and the only active power consumption comes from the RIS controller, allowing low-power and energy-efficient implementations that do not incorporate additional RF radiation into the environment.  However, the potential performance gains may be limited in practice if the number of reflective elements is not sufficiently large. This has motivated the research on active RIS alternatives \cite{Long.2021}, where the reflecting elements incorporate reflection-type amplifiers implemented without high-cost and power-hungry RF chains. Such active circuitry allows simultaneously altering the phase and amplitude to enhance the signal power at the receiver. Nevertheless, they incur modestly higher power consumption and require more evolved CSI acquisition procedures. Identifying passive vs active RISs performance trade-offs in the context of RF-WPT, where noise power is not an issue, while including overhead and protocol issues, are open research directions.


Meanwhile, LISs are low-power/cost massive MIMO metasurfaces. They are equipped with RF circuits and signal processing units and composed of a virtually infinite number of elements to form a spatially continuous transceiver aperture. 
LISs can be realized by using an RIS illuminated by a collocated digital beamforming-based transmitter \cite{Jamali.2021}.
The reconfigurability of the RIS provides extra degrees of freedom in this implementation, allowing it to scale the number of passive elements without requiring a larger number of active antennas.
Meanwhile, DMA, another LIS implementation, is comprised of multiple waveguides, e.g. microstrip, and each may embed a large set of radiating metamaterial elements whose frequency response can be externally and individually adjusted~\cite{Azarbahram.2023}. One or several RF chains can feed the waveguides, while the input signals are radiated by all the metamaterial elements. 

\begin{figure}[t!]
    \centering    
    \includegraphics[width=\linewidth]{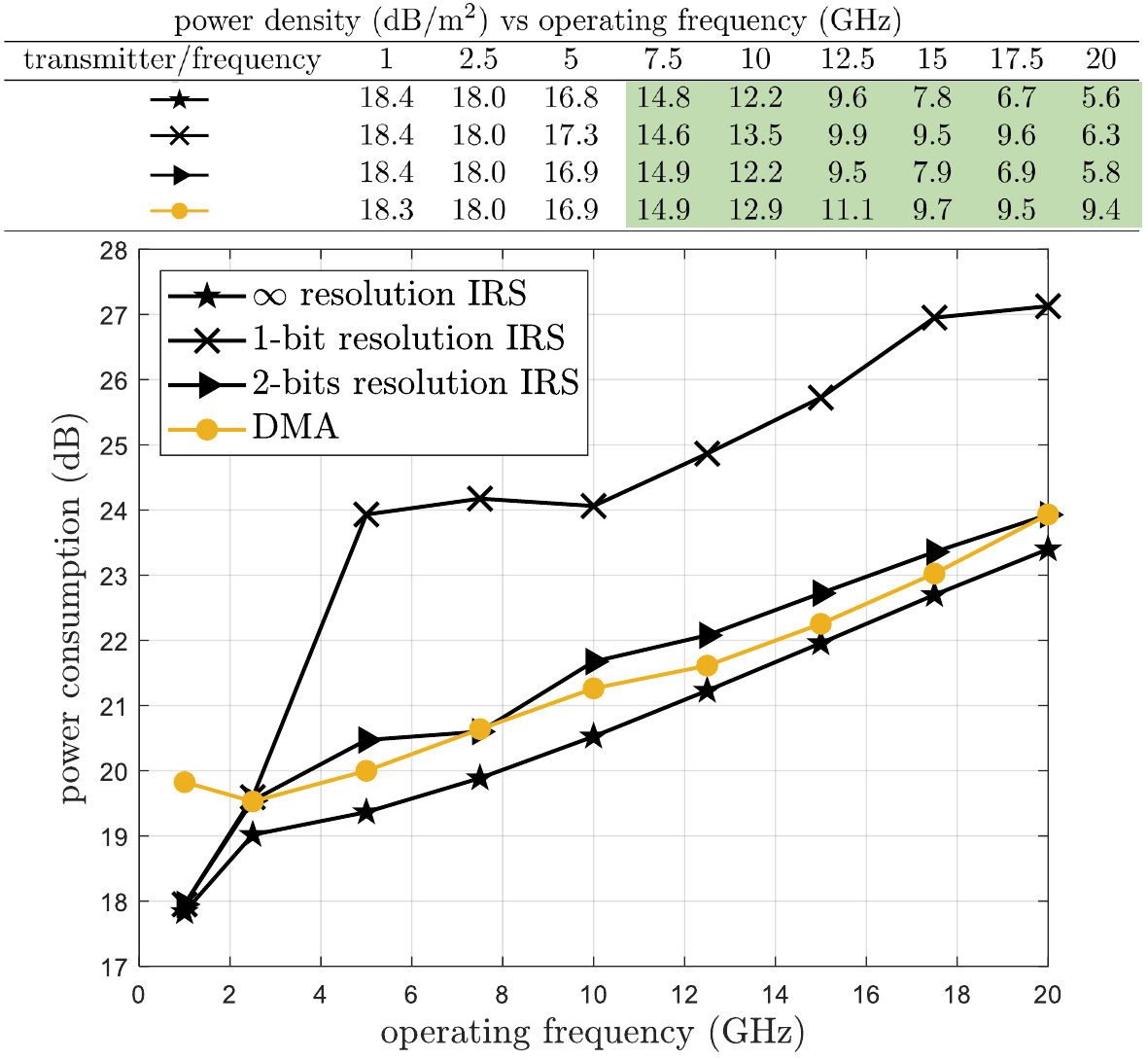}
    \caption{Power consumed by an ET to deliver 1 W of RF power to a single ER located 3 m away in its boresight direction and the resulting power density at 15 cm from the ER as a function of the operation frequency while considering the control board, driving circuits, and HPA's power requirements. The antenna array is square-shaped with a length $L = 50$ cm and comprised of antenna elements with a boresight gain of 7 dB and 13 dB for the RIS- and DMA-based ET implementations, respectively. 
    The ET is equipped with a single RF chain and drives an HPA with a power efficiency of $35\%$. RIS- and DMA-based ET implementations are considered. The RIS-based ET consists of $\left(\lfloor 5L/\lambda \rfloor + 1\right)^2$ passive RIS elements, which are evenly separated in both horizontal and vertical directions by $\lambda/5$. A single antenna feeder, which has a boresight gain of 3 dB, is aligned with the center of the RIS and placed at a distance $4L/\sqrt{\pi}$. In addition, the RIS is equipped with a 1 W control board and driving circuits which consume 5~mW per RIS element to adjust their phase shift. 
    Meanwhile, the DMA-based ET consists of a power splitter, which divides the output power of the RF chain equally among $\lfloor{2L/\lambda}\rfloor+1$ waveguides, each with $\lfloor{5L/\lambda}\rfloor+1$ metamaterial elements. For finite-resolution RIS- and DMA-based scenarios, a particle swarm optimization-based solver is utilized aiming to minimize the power consumption. For the RIS-based ET with infinite phase shift resolution, a simple conjugate phase shifting (concerning the channel coefficients) is optimal and thus adopted here. 
    We highlight the power density values complying with the EMF regulations  listed in Table~\ref{tab:EMF}.}
    \label{fig:result2}
\end{figure}

Fig.~\ref{fig:result2} illustrates the power consumption of RIS and DMA based ETs as a function of frequency for a single-ER setup and ET form factor. Herein, we also plot the resulting power density at 15 cm from the ER. Observe that both architectures exhibit a similar performance, especially as the frequency increases. Still, a RIS-based ET seems preferable at relatively low frequency, although this architecture may be more costly than DMA.
Nevertheless, all this ultimately depends on the scenario geometry and the number of ERs, which are simplified herein and require further investigation. Interestingly, the performance of RIS-based ETs with only two control bits per element is already close to the optimal, obtained from assuming infinite-resolution RIS. Meanwhile, it is observed that given a fixed form factor and charging distance, increasing the operation frequency is appealing for satisfying the EMF radiation exposure constraints. Indeed, ICNIRP regulations are met herein for frequencies above 7.5 GHz in both RIS and DMA based setups.
A critical LIS optimization challenge is encountered as the frequency increases, since the number of variables, i.e., the phase shifts configuration of the antenna elements, becomes massive, for which conventional optimization solvers are inefficient. Two aggravating factors are that: i) in the case of RIS, the phase shifts are discrete; ii) in the case of DMA, more antenna elements can be equipped given a certain form factor and they have a Lorentzian-form response \cite{Azarbahram.2023}. Based on the results illustrated here and after efficiently dealing with the previous challenges, one can speculate that end-to-end PTEs in the order of 10\% may be realized when using greater form-factor LIS based ETs.

Finally, the most investigated metasurfaces are those for passive beamforming/reflection applications. Other metamaterial functions such as refraction, absorption, collimation, polarization change, splitting, and analog processing can be software-engineered using accurate physics-compliant models to promote more flexible and dynamic metasurface-based WPT implementations.
Moreover, one would ideally distribute the WPT antennas and metasurfaces over the entire charging area, what is not aesthetically or economically attractive in most cases. It is more appealing favoring imperceptible installation via hybrid deployments of distributed and centralized antenna arrays, e.g., using radio stripes \cite{Lopez.2022}, together with scenario-specific metasurface assemblage. The use of hardware-flexible ETs such as motor-equipped and flying ETs may be attractive.

\subsection{Harvesters}
The ER circuit blocks must be properly designed for high  RF power input levels and efficiency at the desired frequencies. Multi-stage rectifiers using Schottky diodes (due to their low forward voltage drop and fast switching characteristics) are attractive for this. The higher the incident RF power level targets, the greater the number of rectifying stages/diodes required.  
In such a large-scale signal regime, diodes'
behavior is predominantly resistive, which may facilitate operation optimization with respect to the small-scale signal regime where diodes predominantly behave non-linearly.
Meanwhile, capacitors with low equivalent series resistance and high voltage ratings should be selected as they i) incur lower power losses and can handle higher currents more efficiently without the risk of damage; ii) help maintain a more stable output voltage by reducing voltage droop or ripple caused by internal resistance; and iii) respond quickly to changes in voltage, allowing for faster charging cycles in the charge pump circuit. Nevertheless, such capacitors may usually induce higher parasitic effects which must be taken care of. Other two fundamental ways to increase the harvestable energy are increasing: i) the number of receive antennas, which requires either DC, RF, or hybrid combining mechanisms;\footnote{In Fig.~\ref{fig:system}, we illustrate where RF and DC combining circuitry may be located within an ER. Notice that DC combining is purely passive, while RF combining may demand some active circuitry to coherently combine the RF signals. Hybrid mechanisms trade off the pros and cons for pure DC and RF combining architectures, and are usually preferred, especially when the number of receive antennas is relatively large.} and ii) the frequency spectrum utilization via wideband or multi-band transmissions. 

The ER design can be a complex task, especially when weighing form factor and cost, requiring simulation, prototyping, and iterative optimization. Increasing the antenna's effective size may not only incur higher cost and complexity but may also increase the local area of  EMF radiation exposure above the safety limits (see Fig.~\ref{fig:result1}a) if the RF charging gains do not scale proportionally. Interestingly, metamaterials-based ERs constitute a promising research direction for realizing ultra-compact and high-efficiency ERs (see \cite{Zhou.2022} and references therein). Small form factors can be realized due to the sub-wavelength periodicity of the unit cells within the lattice structure, while a wide-beamwidth and polarization-independent operation together with built-in matching networks and intrinsically high antenna gains promote high efficiency. Nevertheless, this inherently supposes separating WPT and wireless information transfer (WIT) functions, which may not be always appealing.

\subsection{Auxiliary Nodes}\label{ANs}

Auxiliary nodes such as thermal sensors, light detection and ranging (LIDAR), capacitive proximity sensors, radars, and cameras can assist in sensing and locating target ERs, living and non-living obstacles. 
Sensing and positioning services can efficiently support EMF-compliant exposure guarantees, by detecting the presence of living species and accordingly tuning (or even disabling) the high-power RF-WPT service. 

Although the ETs deployment alone or with a coexistent communication infrastructure may support sensing and positioning services, via basic integrated sensing and communication (ISAC) mechanisms, additional accuracy guarantees are desired. These can be provided by a low-complexity sensing infrastructure, especially for the challenging cases where target living species do not carry or wear any device, via device-free localization/sensing \cite{Denis.2019}. 
The degenerative impact of imperfect sensing information on EMF radiation exposure must be considered while providing risk-aware guarantees.
Finally, such auxiliary nodes could incorporate RF sensing capabilities that could help to establish RF energy maps that allow more efficient control of the charging services and EMF exposure in the target area.

\begin{figure}[t!]
    \centering
    \includegraphics[width=0.95\linewidth]{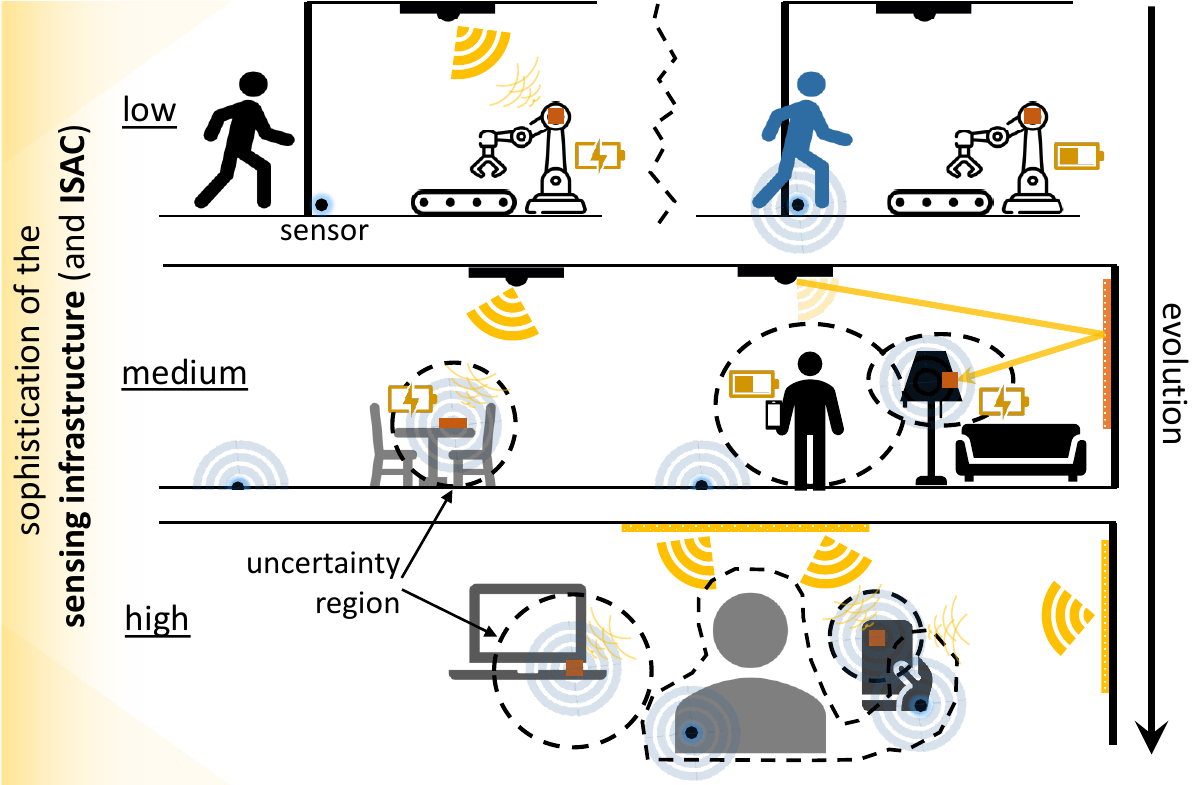}
    \caption{Illustration of different sophistication levels of the sensing infrastructure and charging approaches.}
    \label{sensingProtocol}
\end{figure}

\subsection{Charging Protocols and Management System}\label{control}


The charging protocols include WPT and WIT signaling and rely on information provided by the control plane. WIT signaling among ETs and ERs is required for CSI acquisition, charging requests, accurate positioning/sensing, and mobile edge computing (MEC) such that the devices can fully or partially offload heavy computation tasks. MEC optimization should shift from minimizing devices' energy consumption
to favor the effective system charging efficiency. 


Increasing the system PTE leads to less energy lost in the free space, decreasing the RF contamination and EMF exposure. Still, EMF-friendly protocols, EMF-exposure measurement reports, and EMF-compliant exposure guarantees are needed.
As illustrated in Fig.~\ref{sensingProtocol}, the specific protocols depend on the scenario and the sophistication of the sensing infrastructure, which may be i) ``low'' when it can detect the presence or not of a human/animal in a target area; ii) ``medium'' when it can determine the position of the ERs and closest living species with limited accuracy; or iii) ``high'' when it can do that with high accuracy (e.g., cm-accuracy) and may even measure exposure data with wearables and/or sensors in the proximity of humans. 

In the case of low-sophistication sensing infrastructure, high-power charging services may be allowed only when the target area is human/animal-free. A use case may be an industrial facility for which the RF charging service is deactivated when a human is detected. In the case of medium/high-sophistication sensing infrastructure, high-power charging services may be allowed even in the presence of humans as long as the estimated EMF exposure is safely limited. More sophisticated sensing infrastructures demand more advanced ISAC systems.

%

%

Charging requests may need to be secured and validated, e.g., via distributed ledgers, for micro-billing in public spaces, while forecasting charging demands may be more efficient in massive industrial IoT setups. Proper charging scheduling may benefit significantly from accurate predictions of future demands. 
Scheduling several users for simultaneous charging is advantageous in terms of efficiently exploiting the broadcast nature of the wireless medium, but may affect the performance of some ET blocks such as the HPAs (if driven close to saturation) and increase or make it difficult to control the EMF  exposure \cite{Lopez.2022}. Flexible hybrid protocols considering the required charging time, e.g., due to circuit-related constraints of RF harvesters and/or application demands, are needed.

The WPT management system, comprising a local and global component, decides the charging protocols. The local management system (LM) interfaces with the physical WPT hardware while performing moderate-to-low complexity tasks such as CSI processing, charging requests/forecasts processing, EMF exposure estimation, edge computing, and small-scale optimization and learning from limited, arbitrarily structured, and/or noisy data. 
Taming time-related dynamics such as the number of devices/humans in the loop, users in motion, and time-varying energy demands and computation requirements is essential.
The LM should be interfaced with the global management system (GM) component via QoS-constrained cloud links. GM manages large-scale optimization tasks, including evolved artificial intelligence (AI) mostly over long-term data, exploiting digital twins as testbeds. GM is needed to keep the WPT hardware in the customer's facility the simplest and cheapest possible. However, although the GM subsystem may provide substantial optimization gains, it is not strictly required. The envisioned RF charging systems may be seen in a 6G context as in-X subnetworks \cite{Nishant.2022}, specifically in-house/room/plant RF-WPT subnetworks according to the application. 
The WPT system may either co-exist (separate functioning with dedicated hardware infrastructure), cooperate (the LM and WPT are directly interfaced with the connectivity local service provider), or be co-designed (joint functioning with the same hardware and LM and GM support) with the wireless information network infrastructure while providing added value. The third implementation option may be the final evolution step of a functional WPT system.

\section{Conclusion}
	%
We presented our vision of a cyber-physical system for efficient and safe high-power RF charging, 
while we overviewed the main factors affecting the end-to-end PTE, EMF exposure metrics, and safety limits. 
We showed that the  EMF exposure level close to the EH device can be more easily controlled when operating in near-field conditions. Also, and unexpectedly, the EMF exposure can be more easily controlled at relatively high (low) frequencies when operating in far (near)-field conditions given a fixed number of transmit antennas. 
Contrarily, given form factor constraints such that the number of antennas scales with the operation frequency, the EMF radiation exposure can be more easily controlled at high frequencies. 
Finally, we highlighted the need for high end-to-end PTE architectures and charging protocols transparently complying with EMF exposure regulations.



	\bibliographystyle{IEEEtran}
	\bibliography{IEEEabrv,references}

\begin{thebibliography}{10}
\providecommand{\url}[1]{#1}
\csname url@samestyle\endcsname
\providecommand{\newblock}{\relax}
\providecommand{\bibinfo}[2]{#2}
\providecommand{\BIBentrySTDinterwordspacing}{\spaceskip=0pt\relax}
\providecommand{\BIBentryALTinterwordstretchfactor}{4}
\providecommand{\BIBentryALTinterwordspacing}{\spaceskip=\fontdimen2\font plus
\BIBentryALTinterwordstretchfactor\fontdimen3\font minus
  \fontdimen4\font\relax}
\providecommand{\BIBforeignlanguage}[2]{{%
\expandafter\ifx\csname l@#1\endcsname\relax
\typeout{** WARNING: IEEEtran.bst: No hyphenation pattern has been}%
\typeout{** loaded for the language `#1'. Using the pattern for}%
\typeout{** the default language instead.}%
\else
\language=\csname l@#1\endcsname
\fi
#2}}
\providecommand{\BIBdecl}{\relax}
\BIBdecl

\bibitem{Lopez.2021}
O.~L.~A. L\'opez, H.~Alves, R.~D. Souza, S.~Montejo-S\'anchez, E.~M.~G.
  Fern\'andez, and M.~Latva-Aho, ``Massive wireless energy transfer: Enabling
  sustainable {IoT} toward {6G} era,'' \emph{IEEE Internet of Things Journal},
  vol.~8, no.~11, pp. 8816--8835, 2021.

\bibitem{Ginting.2020}
L.~Ginting, H.~S. Yoon, D.~I. Kim, and K.~W. Choi, ``Beam avoidance for human
  safety in radiative wireless power transfer,'' \emph{IEEE Access}, vol.~8,
  pp. 217\,510--217\,525, 2020.

\bibitem{Clerckx.2021}
B.~Clerckx, K.~Huang, L.~R. Varshney, S.~Ulukus, and M.-S. Alouini, ``Wireless
  power transfer for future networks: Signal processing, machine learning,
  computing, and sensing,'' \emph{IEEE Journal of Selected Topics in Signal
  Processing}, vol.~15, no.~5, pp. 1060--1094, 2021.

\bibitem{Lopez.2022}
O.~L.~A. L\'opez, D.~Kumar, R.~D. Souza, P.~Popovski, A.~T\"olli, and
  M.~Latva-aho, ``Massive {MIMO} with radio stripes for indoor wireless energy
  transfer,'' \emph{IEEE Transactions on Wireless Communications}, pp. 1--1,
  2022.

\bibitem{Zhang.2022}
H.~Zhang, N.~Shlezinger, F.~Guidi, D.~Dardari, M.~F. Imani, and Y.~C. Eldar,
  ``Near-field wireless power transfer for {6G} internet of everything mobile
  networks: Opportunities and challenges,'' \emph{IEEE Communications
  Magazine}, vol.~60, no.~3, pp. 12--18, 2022.

\bibitem{Azarbahram.2023}
A.~Azarbahram, O.~L.~A. Lopez, B.~Clerckx, and M.~Latva-Aho, ``Waveform and
  beamforming optimization for wireless power transfer with dynamic metasurface
  antennas,'' \emph{arXiv preprint arXiv:2307.01081}, 2023.

\bibitem{Zhou.2022}
J.~Zhou, P.~Zhang, J.~Han, L.~Li, and Y.~Huang, ``Metamaterials and
  metasurfaces for wireless power transfer and energy harvesting,''
  \emph{Proceedings of the IEEE}, vol. 110, no.~1, pp. 31--55, 2022.

\bibitem{iarcclassification}
\BIBentryALTinterwordspacing
\emph{{IAR}C Monographs on the Evaluation of Carcinogenic Risks to
  Humans}.\hskip 1em plus 0.5em minus 0.4em\relax Lyon, France: International
  Agency for Research on Cancer, 2021. [Online]. Available:
  \url{monographs.iarc.who.int}
\BIBentrySTDinterwordspacing

\bibitem{ICNIRP.2020}
I.~C. on~Non-Ionizing Radiation~Protection \emph{et~al.}, ``Guidelines for
  limiting exposure to electromagnetic fields (100 {kHz to 300 GHz)},''
  \emph{Health physics}, vol. 118, no.~5, pp. 483--524, 2020.

\bibitem{Lopez.2023}
O.~A. L{\'o}pez, O.~M. Rosabal, D.~Ruiz-Guirola, P.~Raghuwanshi, K.~Mikhaylov,
  L.~Lov{\'e}n, and S.~Iyer, ``Energy-sustainable {IoT} connectivity: Vision,
  technological enablers, challenges, and future directions,'' \emph{arXiv
  preprint arXiv:2306.02444}, 2023.

\bibitem{Long.2021}
R.~Long, Y.-C. Liang, Y.~Pei, and E.~G. Larsson, ``Active reconfigurable
  intelligent surface-aided wireless communications,'' \emph{IEEE Transactions
  on Wireless Communications}, vol.~20, no.~8, pp. 4962--4975, 2021.

\bibitem{Jamali.2021}
V.~Jamali, A.~M. Tulino, G.~Fischer, R.~R. M\"{u}ller, and R.~Schober,
  ``Intelligent surface-aided transmitter architectures for millimeter-wave
  ultra massive {MIMO} systems,'' \emph{IEEE Open Journal of the Communications
  Society}, vol.~2, pp. 144--167, 2021.

\bibitem{Denis.2019}
S.~Denis, R.~Berkvens, and M.~Weyn, ``A survey on detection, tracking and
  identification in radio frequency-based device-free localization,''
  \emph{Sensors}, vol.~19, no.~23, p. 5329, 2019.

\bibitem{Nishant.2022}
N.~Batra, H.~Holma, H.~Viswanathan, T.~Wild, P.~Baracca, G.~Berardinelli,
  G.~Kunzmann, V.~Ziegler, M.~Montag, P.~Merz, and P.~{Vetter (Editors)},
  ``Envisioning a {6G} future,'' Nokia Bell Labs, e-Book, 2022.

\end{thebibliography}
	%



\end{document}